# Action in a fractal universe and the holographic upper bound

Scott Funkhouser
Department of Physics, The Citadel
171 Moultrie St., Charleston, SC 29409

ABSTRACT
The basic scaling laws for structures in a fractal universe require that the characteristic quantity of action associated with astronomical bodies should be of order near the maximum possible action allowed by the holographic upper bound. That conclusion is consistent with the observed parameters of galaxies and clusters.

If the predominant structures of the universe are arranged in a self-similar manner then the parameters of bodies on any given structural level should be characterized well by at least two, principal scaling laws. Consider some structural level $i$ within a fractal universe that is populated by bodies with a characteristic mass $M_i$ that is contained with a spherical cell whose radius is $R_i$. Let each body on structural level $i$ contain at least one subordinate level of fractal structure, denoted by $j$ that is populated by bodies whose characteristic masses are $M_j$ and that are contained within a characteristic cell of radius $R_j$. According to the basic requirements of fractal structure, it should be that

$$\frac{M_i}{R_i^D} \sim \frac{M_j}{R_j^D}, \qquad (1)$$

where $D$ is the fractal dimension relating the two levels [1]. A second, primary scaling law associated with fractal structures involves the characteristic quantity of action that is associated with the bodies on each level of structure. The action $A_i$ associated with each body on structural level $i$ must be related to the action $A_j$ according to [1]

$$A_i \sim A_j \left(\frac{M_i}{M_j}\right)^{(D+1)/D}. \qquad (2)$$

With a substitution from (1), Eq. (2) may be expressed as

$$A_i \sim A_j \frac{M_i R_i}{M_j R_j}. \qquad (3)$$

The characteristic quantity of action for gravitationally virialized systems is roughly equal to the product of the gravitational potential energy of the body and the associated relaxation time [1]. The gravitational potential energy of a body of mass $M$ contained within a sphere of radius $R$ is $GM^2/R$. The corresponding relaxation time is roughly $(R^3 G^{-1} M^{-1})^{1/2}$, which is also the inverse of the characteristic, virialized rotational frequency. The action $A$ of a gravitationally virialized body is thus of the order $(GM^3 R)^{1/2}$, which is also roughly the characteristic angular momentum of the system.

If the fractal nature of the universe extends to the realm of particles then the relationship in (3) should apply to the parameters of any given astronomical body in relation to the fundamental particles that constitute the majority of the mass of the body. Let there be some fundamental particle whose mass is $m$ that constitutes the preponderance of the mass of any given astronomical body. That particle may be the nucleon, but the analysis presented here should be relevant even if the masses of cosmic structures are dominated by some yet-unidentified particle. The characteristic action and angular momentum of any fundamental particle must be of the order the Planck quantum

$\hbar$. The characteristic cell-radius of the fundamental particle should be of order near the Compton wavelength $\hbar/(mc)$ [1]. Therefore, if the level $j$ of fractal structure in (1) – (3) should represent the fundamental particle that constitutes the majority of the masses of the bodies on structural level $i$ then it must be that

$$A_i \sim M_i R_i c. \qquad (4)$$

The expression for the characteristic action in (4) is significant in the context of the thermodynamics of black holes and holographic principles. The maximum possible angular momentum allowed to any body whose mass $M$ is contained within a sphere of radius $R$ is $MRc$, where $c$ is the vacuum-speed of light [2]. Thus, the quantities of action associated with bodies in a fractal universe should be near the maximum possible action allowed to them. The expression in (4) suggests also that the relationship among action, mass and cell-radius that characterizes particles is also relevant for macroscopic bodies in a fractal universe.

The mass of a typical galaxy is of the order $10^{42}$kg and the characteristic galactic cell-radius is of the order $10^{20}$m. The characteristic action of galaxies is therefore roughly $10^{69}$Js. The holographic upper bound on the action or angular momentum of a typical galaxy is of the order $10^{71}$Js. Thus, within a few orders of magnitude, which is a relatively small discrepancy, the parameters of galaxies are consistent with (4). The mass of a typical cluster of galaxies is of the order $10^{45}$kg, and the characteristic radius of a typical cluster is of the order $10^{22}$m. The action of a typical cluster of galaxies should be consequently of the order $10^{74}$Js. The upper bound on the action of a cluster of galaxies is of the order $10^{76}$Js. The parameters of clusters of galaxies are therefore also consistent with the scaling law in (4). Note also that the cell-radius associated with fundamental particles may not be exactly equal to the Compton wavelength, thus introducing some additional uncertainty.